\documentclass[12pt]{article}

\usepackage{array,dsfont} 
\usepackage{epsfig}
\usepackage{amssymb}
\usepackage{graphics,graphpap}

\renewcommand{\thefootnote}{\fnsymbol{footnote}}

\begin{document}

\begin{titlepage}
\begin{flushright}\begin{tabular}{l}
IPPP/07/10\\
DCPT/07/20
\end{tabular}
\end{flushright}
\vskip1.5cm
\begin{center}
   {\Large \bf\boldmath Implication of the $D^0$ Width Difference\\[5pt] 
    On CP-Violation in $D^0$-$\bar D^0$ Mixing} 
    \vskip2.5cm {\sc
Patricia Ball\footnote{Patricia.Ball@durham.ac.uk}
}
  \vskip0.5cm
{\em         IPPP, Department of Physics,
University of Durham, Durham DH1 3LE, UK }\\
\vskip2.5cm 


\vskip4cm

{\large\bf Abstract\\[10pt]} \parbox[t]{\textwidth}{
Both BaBar and Belle have found evidence for a non-zero width
difference in  the $D^0$-$\bar D^0$ system. Although there is no direct
experimental evidence for CP-violation in $D$ mixing (yet), we show
that the measured values of the width difference $y\sim \Delta\Gamma$
already imply constraints on the CP-odd phase in $D$ mixing,
which, if significantly different from zero, would be an unambiguous
signal of new physics.
}

\end{center}
\end{titlepage}

\setcounter{footnote}{0}
\renewcommand{\thefootnote}{\arabic{footnote}}

\newpage

The highlight of this year's Moriond conference on electroweak 
interactions and unified theories arguably was the announcement by
BaBar and Belle of experimental evidence for $D^0$-$\bar D^0$ mixing
\cite{talks,BaBar,Belle}, which was quickly followed by a number of theoretical
analyses \cite{fast,nir,me,buras,he,8}. While Refs.~\cite{fast,buras,he,8}
focused on the constraints posed, by the experimental results, on
various new-physics models, Ref.~\cite{nir} presented a first analysis
of the implications of these results for the fundamental
parameters describing $D$ mixing. The
purpose of this letter is to show that the present experimental results
already imply 
constraints on a sizeable CP-odd phase in $D$ mixing, which could only be
due to new physics (NP).

To start with, let us shortly review the theoretical
formalism of $D$ mixing and the experimental results, see
Refs.~\cite{review1,review2} for more detailed reviews.
In complete analogy to $B$ mixing, $D$ mixing in the SM is due to box
diagrams with internal quarks and $W$ bosons. In contrast to $B$,
though, the internal quarks are down-type. Also in contrast to $B$
mixing, the GIM mechanism is much more effective, as the contribution
of the heaviest down-type
quark, the $b$, comes with a relative enhancement factor
$(m_b^2-m_{s,d}^2)/(m_s^2-m_d^2)$, but also a large CKM-suppression
factor $|V_{ub}^{\phantom{*}}V_{cb}^*|^2/|V_{us}^{\phantom{*}} 
V_{cs}^*|^2\sim \lambda^8$, which
renders its contribution to $D$ mixing $\sim 1\%$ and hence negligible.
As a consequence, $D$ mixing is 
very sensitive to the potential intervention of NP. On
the other hand, it is also rather difficult to calculate the
SM ``background'' to $D$ mixing, 
as the loop-diagrams are dominated by $s$ and $d$ quarks
and hence sensitive to the intervention of resonances and
non-perturbative QCD. 
The quasi-decoupling of the
3rd quark generation also implies that CP violation in $D$ mixing is
extremely small in the SM, and hence any observation of CP violation
will be an unambiguous signal of new physics, independently of hadronic
uncertainties. 

The theoretical parameters describing $D$ mixing can be defined in
complete analogy to those for $B$ mixing: the time evolution of the
$D^0$ system is described by the Schr\"odinger equation
\begin{equation}
i\frac{\partial}{\partial t}\left(\begin{array}{c} D^0(t) \\ \bar
    D^0(t)\end{array}
\right) =  \left(M - i\, \frac{\Gamma}{2}\right)
\left(\begin{array}{c} D^0(t) \\ \bar
    D^0(t)\end{array}
\right)
\end{equation}
with Hermitian  matrices $M$ and $\Gamma$. The off-diagonal elements
of these matrices, $M_{12}$ and $\Gamma_{12}$, describe, respectively, 
the dispersive and absorptive parts of $D$ mixing.
The flavour-eigenstates
$D^0=(c\bar u)$, $\bar D^0=(u\bar c)$ are related to the mass-eigenstates
$D_{1,2}$ by
\begin{equation}\label{2}
|D_{1,2}\rangle = p |D^0\rangle \pm q | \bar D^0\rangle
\end{equation}
with
\begin{equation}\label{3}
\left( \frac{q}{p}\right)^2 =
\frac{M_{12}^*-\frac{i}{2}\,\Gamma^*_{12}}{ 
M_{12}-\frac{i}{2}\,\Gamma_{12}}\,;
\end{equation}
$|p|^2+|q|^2=1$ by definition. 

The basic observables in $D$ mixing are the mass and lifetime
difference of $D_{1,2}$, which are usually normalised to the
average lifetime $\Gamma = (\Gamma_1+\Gamma_2)/2$:
\begin{equation}
x\equiv \frac{\Delta M}{\Gamma} = \frac{M_2-M_1}{\Gamma}\,,\quad
y\equiv \frac{\Delta \Gamma}{2\Gamma} =
\frac{\Gamma_2-\Gamma_1}{2\Gamma}\,.
\end{equation}
In this letter we follow the sign convention of Ref.~\cite{nir}, according
to which $x$ is positive by definition. The sign of $y$ then has to be
determined from experiment.
In addition, if there is CP-violation in the $D$ system, one also has 
\begin{equation}\label{5}
\left|\frac{q}{p}\right| \neq 1,\quad \phi\equiv
{\rm arg}(M_{12}/\Gamma_{12})\neq 0.
\end{equation} 

While previously only bounds on $x$ and $y$ were known, both BaBar and Belle
have now found evidence for non-vanishing mixing in the $D$
system. BaBar has obtained this evidence from the measurement of the
doubly Cabibbo-suppressed decay $D^0\to K^+\pi^-$ (and its CP
conjugate), yielding
\begin{eqnarray}
y' &=& (0.97\pm 0.44({\rm stat})\pm 0.31({\rm syst}))
\times 10^{-2},\nonumber\\ 
x'^2 &=& (-0.022\pm 0.030({\rm stat})\pm 0.021({\rm syst}))\times
10^{-2},
\end{eqnarray}
while Belle obtains
\begin{equation}\label{8a}
y_{\rm CP} = (1.31\pm 0.32({\rm stat})\pm 0.25({\rm syst}))\times
10^{-2}
\end{equation}
from $D^0\to K^+K^-, \pi^+\pi^-$ and 
\begin{equation}
x = (0.80\pm 0.29({\rm stat})\pm 0.17({\rm syst}))\times 10^{-2},\quad
y = (0.33\pm 0.24({\rm stat})\pm 0.15({\rm syst}))\times 10^{-2}
\end{equation}
from a Dalitz-plot analysis of $D^0\to K_S^0\pi^+\pi^-$.
Here $y_{\rm CP}\to y$ in the limit of no CP violation in $D$ mixing,
while the primed quantities $x',y'$ are related to $x,y$ by a rotation
by a strong phase $\delta_{K\pi}$:
\begin{equation}
y' = \cos\delta_{K\pi} - x \sin\delta_{K\pi},\quad x' = x\cos\delta_{K\pi}
+ y\sin\delta_{K\pi}.
\end{equation}
Limited experimental information on this
phase has been obtainted at CLEO-c \cite{cos}:
\begin{equation}\label{exp}
\cos \delta_{K\pi} = 1.09\pm 0.66\,,
\end{equation}
which can be translated into $\delta_{K\pi} = (0\pm 65)^\circ$.
An analysis with a larger data-set is underway at CLEO-c, with an expected
uncertainty of $\Delta \cos \delta_{K\pi}\approx 0.1$ in the next
couple of years
\cite{asner}; BES-III is expected to reach $\Delta \cos \delta_{K\pi}\approx
0.04$ after 4 years of running \cite{BES}.
The experimental result (\ref{exp}) agrees with theoretical expectations,
$\delta_{K\pi} = 0$ in the SU(3)-limit and $|\delta_{K\pi}|\,
\raisebox{-4pt}{$\stackrel{<}{\sim}$}\,
 15^\circ$ from a calculation of the amplitudes in QCD
factorisation \cite{gao}.
Based on these experimental results, 
a preliminary HFAG-average 
was presented at the 2007 CERN workshop ``Flavour in the
Era of the LHC'' \cite{asner}:
\begin{equation}\label{HFAG}
x = (8.5^{+3.2}_{-3.1})\times 10^{-3},\quad y =
(7.1^{+2.0}_{-2.3})\times 10^{-3}.
\end{equation}
Adding errors in quadrature, this implies
\begin{equation}\label{new}
\frac{x}{y} = 1.2\pm 0.6.
\end{equation}
The exact relations between $\Delta M$, $\Delta\Gamma$, $M_{12}$ and
$\Gamma_{12}$ are given by
\begin{eqnarray}
(\Delta M)^2 - \frac{1}{4}\, (\Delta\Gamma)^2 & = & 4 |M_{12}|^2 -
  |\Gamma_{12}|^2 ,\nonumber\\
(\Delta M)(\Delta\Gamma) & = & 4 {\rm Re}(M_{12}^*\Gamma_{12}) = 4 |M_{12}|
  |\Gamma_{12}| \cos\phi\,.\label{relations}
\end{eqnarray}
Eq.~(\ref{relations}) implies $x/y>0$ for $|\phi|<\pi/2$ and $x/y<0$
for $\pi/2<|\phi|<3\pi/2$. In view of the above experimental results,
we assume $|\phi|<\pi/2$ from now on. 

As for the CP-violating observables, $|q/p|\neq 1$ characterises
CP-violation in mixing and can be measured for instance 
in flavour-specific decays
$D^0\to f$, where $\bar D^0\to f$ is possible only via mixing. The
prime example is semileptonic decays with
\begin{equation}\label{11}
A_{\rm SL} = \frac{\Gamma(D^0\to \ell^- X) - \Gamma(\bar D^0\to \ell^+
  X)}{\Gamma(D^0\to \ell^- X) + \Gamma(\bar D^0\to \ell^+ X)} =
  \frac{|q/p|^2 - |p/q|^2}{|q/p|^2 + |p/q|^2}\,.
\end{equation}
Although the B factories may have some sensitivity to this asymmetry,
its measurement is severely impaired by the fact that $D$ mixing
proceeds only very slowly, resulting in a large suppression factor of
the mixed vs.\ the unmixed rate:
\begin{equation}
\frac{\Gamma(D^0\to \ell^- X)}{\Gamma(D^0\to \ell^+ X)} =
\frac{x^2+y^2}{2+x^2+y^2} \approx 6\times 10^{-5}.
\end{equation}

\begin{figure}
$$\epsfxsize=0.5\textwidth\epsffile{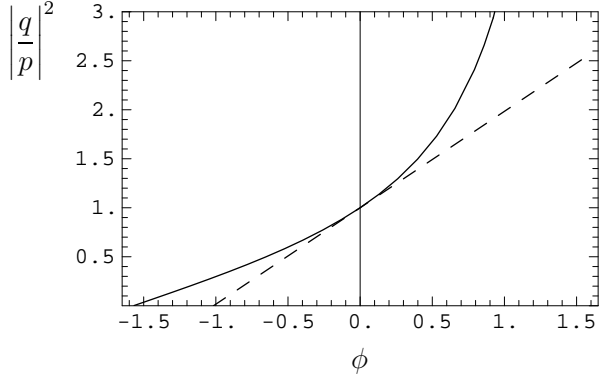}$$
\vskip-15pt
\caption[]{\small $|q/p|^2$, Eq.~(\ref{full}), 
as a function of the CP-odd phase $\phi$
  for the central experimental value $\tilde r = 7.1/8.5$. Solid line:
full expression, dashed line: first order expansion around $\phi=0$.
}\label{fig1}
\end{figure}

Both in the $K$ and the $B$ system the quantity
\begin{equation}
A_M \equiv \left|\frac{q}{p}\right|-1 
\end{equation}
is very small, which
however need not necessarily be the case for $D$'s. From (\ref{3}) one
derives the general expression 
\begin{equation}
\left| \frac{q}{p}\right|^2 = 
\left( \frac{4 + r^2 + 4 r \sin\phi}{4 + r^2 - 4 r
  \sin\phi}\right)^{1/2}
\end{equation}
with $r=|\Gamma_{12}/M_{12}|$ and the weak phase $\phi$ defined in (\ref{5}). 
In the $B$ system, one has $r\ll 1$ (the
current up-to-date numbers are $r\approx 7\times 10^{-3}$ for $B_d$
and $r\approx 5\times 10^{-3}$ for $B_s$ \cite{LN06}), 
so that upon expansion in $r$
\begin{equation}
\left| \frac{q}{p}\right|^2_{B_{d,s}} = 
1 + \left|\frac{\Gamma_{12}}{M_{12}}\right| \sin\phi + O(r^2).
\end{equation}
Note that this formula refers to the definition $\phi={\rm
  arg}(M_{12}/\Gamma_{12})$, which differs by $+\pi$ from the one used in
  Ref.~\cite{LN06}, $\phi={\rm arg}(-M_{12}/\Gamma_{12})$.
For the $K$ system, one finds $r\approx |\Delta\Gamma/\Delta
M|\approx 2$ from experiment, but now
the phase $\phi$ turns out to be small, so that 
\begin{equation}
\left| \frac{q}{p}\right|^2_{K} = 1 + \frac{4r}{4+r^2}\,\phi +
O(\phi^2) \approx 1 + \phi.
\end{equation}
In both cases, $|q/p|\approx 1$ to a very good approximation. In the
$D$ system, however, there is no natural hierarchy $r\ll 1$, and of
course one hopes that NP-effects  induce
$|\phi|\gg 0$. In this case, and because $x$ and
$y$ have been measured, while $|M_{12}|$ and $|\Gamma_{12}|$ are
difficult to calculate, it is convenient to express
$|q/p|_D$ in terms of $x$, $y$, $\phi$, using the exact relations
(\ref{relations}).
From (\ref{3}), and defining $\tilde r = y/x$, we then obtain
\begin{eqnarray}
\left| \frac{q}{p}\right|^2 & = & \frac{1}{\sqrt{2}(1+\tilde r^2)} \, \left\{
2 (1+\tilde r^2)^2 + 16 \tilde r^2 \tan^2\phi \vphantom{\sqrt{\sin^2
    \phi}} \right.\nonumber\\
&& \hspace*{2.4cm}\left. + 8 \tilde r \tan\phi \sec\phi \sqrt{(1+\tilde r^2)^2
  - (1-\tilde r^2)^2 \sin^2\phi}\right\}^{1/2}.\label{full}
\end{eqnarray}
Note that for finite
$xy$ and $\phi=\pm \pi/2$, $|q/p|$ diverges 
because $x y\to 0$ for $\phi\to
\pm \pi/2$ from (\ref{relations}).
In Fig.~\ref{fig1} we plot $|q/p|^2$ as function of $\phi$, for the
central experimental value from HFAG, $\tilde r = 7.1/8.5$,
Eq.~(\ref{HFAG}). 
It is obvious that even for moderate values of $\phi$ the small-$\phi$
expansion is not really reliable. 

What is the currently available experimental information on
CP-violating in $D$ mixing, i.e.\ $|q/p|$
and $\phi$? 
As already mentioned, the semileptonic CP-asymmetry (\ref{11}) has not
been measured yet. What has been measured, though, is the effect of
CP-violation on the time-dependent rates of $D^0\to K^+\pi^-$ and
$\bar D^0\to K^-\pi^+$. The BaBar collaboration has parametrised these
rates as
\begin{eqnarray}
\Gamma(D^0(t)\to K^+\pi^-) & \propto & e^{-\Gamma t} \left[ R_D +
  \sqrt{R_D} y'_+ \Gamma t + \frac{x'^2_++y'^2_+}{4}\,(\Gamma
  t)^2\right],\nonumber\\
\Gamma(\bar D^0(t)\to K^-\pi^+) & \propto & e^{-\Gamma t} \left[ R_D +
  \sqrt{R_D} y'_- \Gamma t + \frac{x'^2_-+y'^2_-}{4}\,(\Gamma
  t)^2\right]\label{18}
\end{eqnarray}
and fit the $D^0$ and $\bar D^0$ samples separately.
They find \cite{BaBar}
\begin{eqnarray}
y'_+ & = & (9.8\pm 6.4({\rm stat})\pm 4.5({\rm syst})) \times
10^{-3},\nonumber\\
y'_- & = & (9.6\pm 6.1({\rm stat})\pm 4.3({\rm syst})) \times
10^{-3}.
\end{eqnarray}
Adding errors in quadrature, this means $y'_+/y'_- = 1.0\pm 1.1$.
BaBar also obtains values for 
$x'^2_\pm$ which we do not quote here, because the
sensitivity to the quadratic term in (\ref{18}) is less than that to
the linear term in $y'_\pm$. $R_D^{1/2}$ is the ratio of the
doubly Cabbibo-suppressed to the Cabibbo-favoured amplitude, 
$R_D^{1/2} = |A(D^0\to K^+\pi^-)/A(D^0\to K^-\pi^+)|$. $\delta_{K\pi}$
is the relative strong phase in the Cabibbo-favoured and suppressed
amplitudes:
\begin{equation}
\frac{A(D^0\to K^+\pi^-)}{A(\bar D^0\to  K^+\pi^-)} = - \sqrt{R_D}
e^{-i\delta_{K\pi}};
\end{equation}
the minus-sign comes from the relative sign between the CKM matrix
elements $V_{cd}$ and $V_{us}$.
In the limit of
no CP-violation in the decay amplitude, one has $|A(D^0\to K^-\pi^+)| =
|A(\bar D^0\to K^+\pi^-)|$, which is expected to be a very good
approximation, in view of the fact that the decay is solely due to a
tree-level amplitude. Then the relation of $y_\pm'$ to $x$, $y$ and 
$\phi$ is given by
\begin{eqnarray}
y'_+ & = & \left|\frac{q}{p}\right| \left\{ (y \cos\delta_{K\pi} - x
\sin\delta_{K\pi}) \cos\phi + (x\cos\delta_{K\pi} + y\sin\delta_{K\pi})
\sin\phi\right\},\nonumber\\
y'_- & = & \left|\frac{p}{q}\right| \left\{ (y \cos\delta_{K\pi} - x
\sin\delta_{K\pi}) \cos\phi - (x\cos\delta_{K\pi} + y\sin\delta_{K\pi})
\sin\phi\right\}.
\end{eqnarray}

\begin{figure}
$$\epsfxsize=0.48\textwidth\epsffile{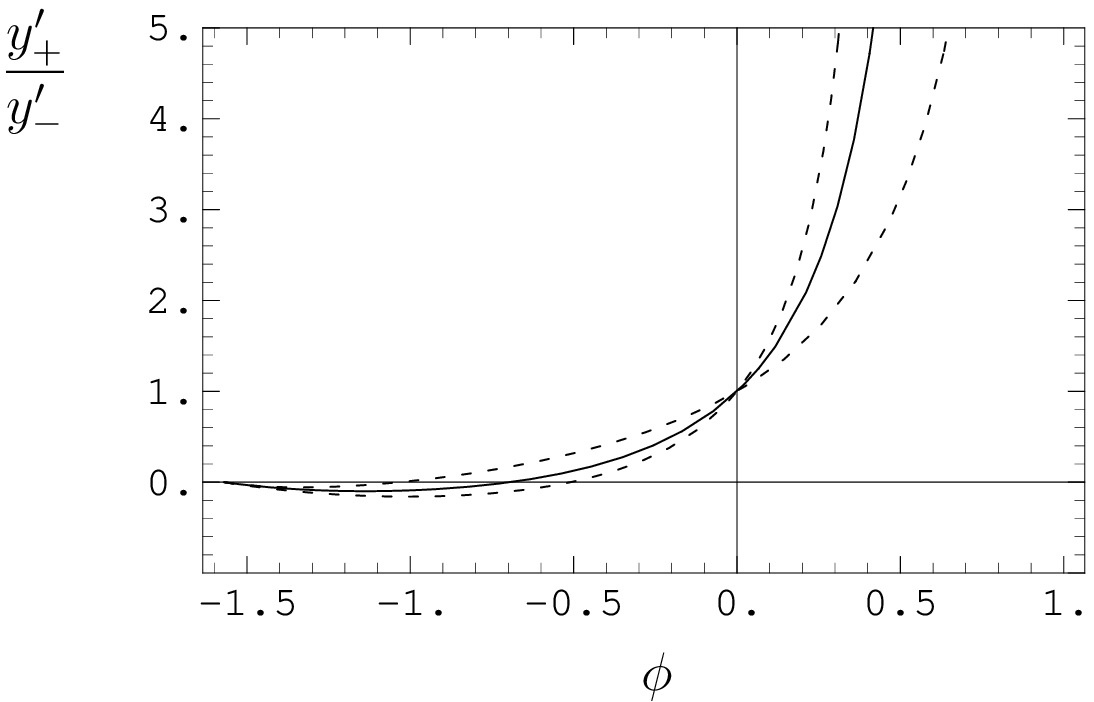}\quad
\epsfxsize=0.48\textwidth\epsffile{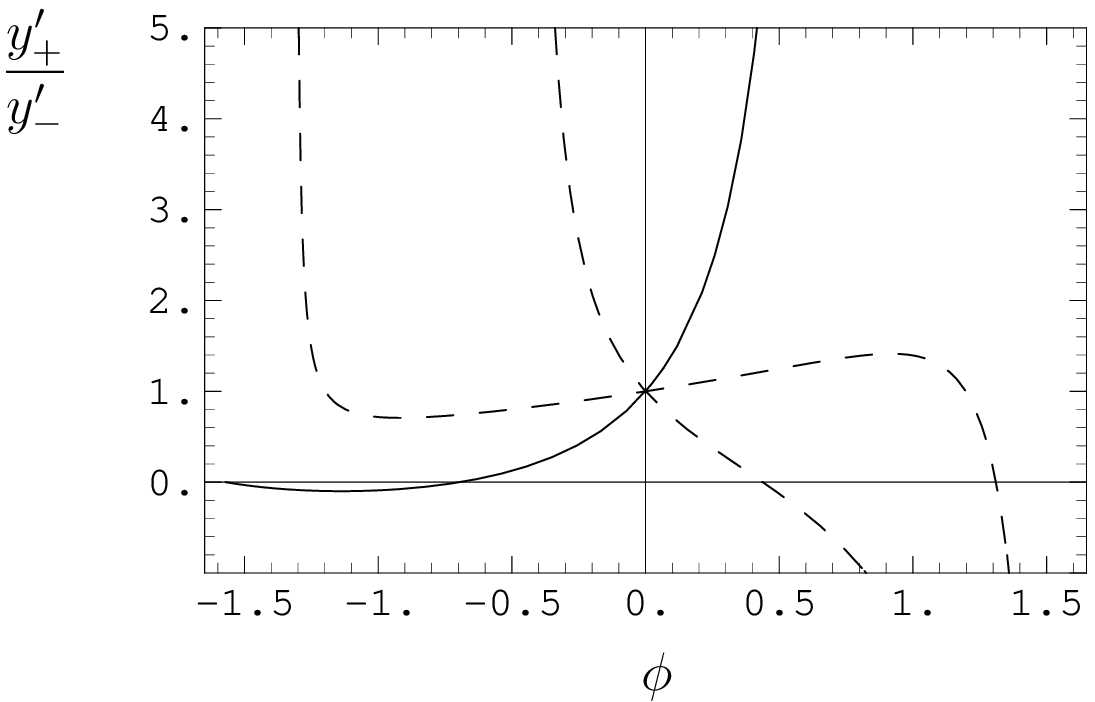}$$
\vskip-20pt
\caption[]{\small Left: $y_+'/y'_-$ as function of $\phi$ for $x/y = 1.2$
  (solid line) and $x/y=\{0.6,1.8\}$ (dashed lines), from 
Eq.~(\ref{HFAG}). $\delta_{K\pi} =
  0$. Right: $y_+'/y'_-$ as function of $\phi$ for $x/y = 1.2$ for
  $\delta_{K\pi} = 0$ (solid line) and $\delta_{K\pi} = \pm 65^\circ$
  (dashed lines).}\label{fig2}
\end{figure}

Presently, the experimental result for  
$y_+'/y'_-$ is compatible with 1, although with
considerable uncertainties. Any significant deviation from 1 would
be a sign for new physics. In Fig.~\ref{fig2} we plot $y'_+/y'_-$ as
function of $\phi$, for different values of $x/y$ and $\delta_{K\pi}$.
The figures clearly show that the value of $y'_+/y'_-$ is very sensitive
to the phase $\phi$, at least if $\delta_{K\pi}$ is not too close to
$-65^\circ$, which corresponds to 
the nearly constant dashed line in Fig.~\ref{fig2}b. The
reason for this dependence on $\delta_{K\pi}$ 
becomes clearer if $y'_+/y'_-$ is expanded to first
order in $\phi$:
\begin{equation}
\frac{y'_+}{y_-} = 1 - 2\phi\,\frac{x(x^2+2y^2) \cos\delta_{K\pi} + y^3
  \sin\delta_{K\pi}}{(x^2+y^2) (x\sin\delta_{K\pi} -
  y\cos\delta_{K\pi})} + O(\phi^2)\,.
\end{equation}
For the central values of $x$ and $y$, Eq.~(\ref{HFAG}), 
this amounts to $1+ 3.4 \phi$ for $\delta_{K\pi} = 0$, $1-3.3\phi$ for
$\delta_{K\pi} = +65^\circ$ and $1+0.45 \phi$  for
$\delta_{K\pi} = -65^\circ$, which explains the shape of the curves in
Fig.~\ref{fig2}b. Evidently it is important to reduce the
uncertainty of $\delta_{K\pi}$, which, as mentioned earlier, will be
achieved within the next few years. On
the other hand, as shown in Fig.~\ref{fig2}a, $y'_+/y'_-$, which 
depends only on the ratio $x/y$, but not $x$ and $y$ separately,
is not very sensitive to the 
precise value of that ratio, but very much so to $\phi$. 
The conclusion is that, even if $x/y$ itself
cannot be determined very precisely, $y_+'/y'_-$
will nonetheless be a powerful tool to constrain $\phi$, at least once
$\delta_{K\pi}$ will be known more precisely. Already now very large
values $\phi\sim \pi/2$ are excluded.

Another, more theory-dependent constraint on $\phi$ can be derived
from the value of $y$. This argument centers around the fact that 
(a) the experimental result
(\ref{HFAG}) is at the top end of theoretical predictions $y_{\rm
  SM}\sim 1\%$ \cite{yth} and (b) new
physics indicated by a non-zero value of  $\phi$ always {\em
  reduces} the lifetime difference, independently of the value of
$x$. This observation is similar to what
was found, some time ago, for the $B_s$ system \cite{yuval}.
In order to derive it, we assume 
that new physics does not affect $\Gamma_{12}$,\footnote{See, however, 
Ref.~\cite{golowich} for a discussion
of the effect of tiny NP admixtures to $\Gamma_{12}$.} so that
$\Gamma_{12} = \Gamma_{12}^{\rm SM}$.
We then have
$2|\Gamma_{12}| = \Delta \Gamma^{\rm SM}$ and hence $|y_{\rm SM}| =
|\Gamma_{12}|/\Gamma$. Using the relations (\ref{relations}), we can
then express the ratio $|\Delta\Gamma/\Delta\Gamma^{\rm SM}|$ in terms
of $y_{\rm SM}$, $x$ and $\phi$:
\begin{equation}\label{eq}
\left|\frac{y}{y_{\rm SM}}\right| 
= \left| \frac{\Delta\Gamma}{\Delta\Gamma^{\rm SM}} \right| = \left(
\frac{y_{\rm SM}^2 + x^2}{y_{\rm SM}^2 + x^2/\cos^2\phi}\right)^{1/2}.
\end{equation}
This implies that {\em new physics
 always reduces the lifetime difference}, independently of the value of
 $x$ (and any new physics in the mass difference). In particular one has
  $y=0$ for $\phi=\pm \pi/2$ and $x\neq 0$, which follows from the 2nd
 relation (\ref{relations}). Eq.~(\ref{eq}) is the manifestation of
  the fact that one does not need to observe CP-violation in order
  to constrain it. A famous example for this is the unitarity triangle 
in $B$ physics, whose sides are determined from CP-conserving 
quantities only, but nonetheless allow a precise measurement of the size of
  CP-violation in the SM, via the angles and the area of the triangle. 
\begin{figure}
$$\epsfxsize=0.5\textwidth\epsffile{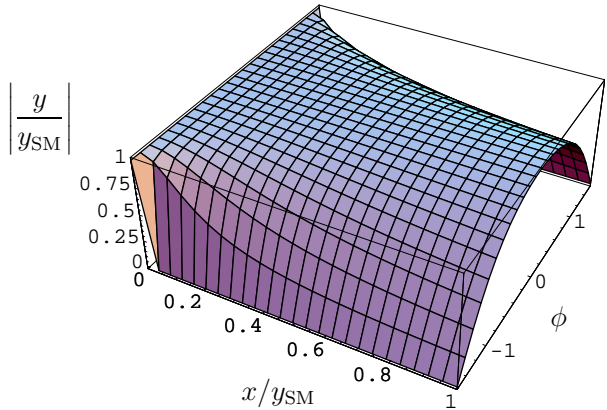}$$
\vskip-15pt
\caption[]{\small Plot of $|\Delta\Gamma/\Delta \Gamma^{\rm SM}|$,
  Eq.~(\ref{eq}), as a function of $x/y_{\rm SM}$ and
  $\phi$.}\label{fig3}
\end{figure}
In Fig.~\ref{fig3}, we plot  $|\Delta\Gamma/\Delta \Gamma^{\rm SM}|$
as a function of $r=x/y_{\rm SM}$. The zero at $\phi=\pm\pi/2$ is
clearly visible. The experimental value $|y/y_{\rm SM}|= O(1)$ then
excludes phases $\phi$ close to $\pm \pi/2$. In order to make more
quantitative statements, apparently a more precise calculation of
$y_{\rm SM}$ is needed.

\begin{figure}
$$\epsfxsize=0.48\textwidth\epsffile{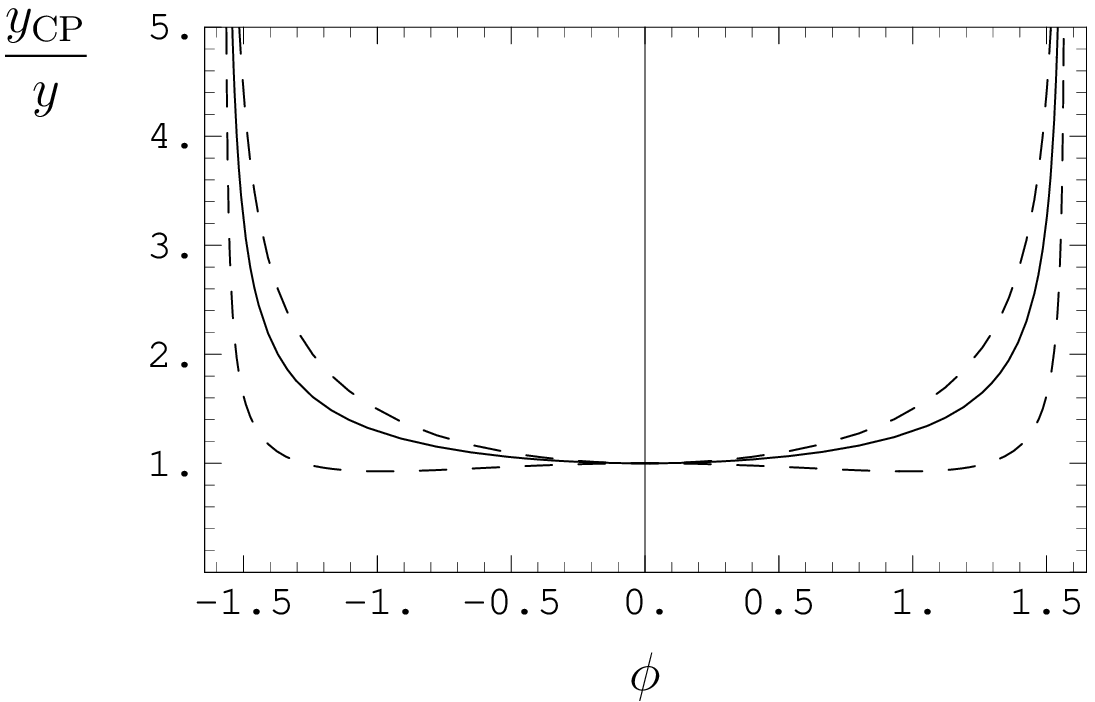}\quad
\epsfxsize=0.48\textwidth\epsffile{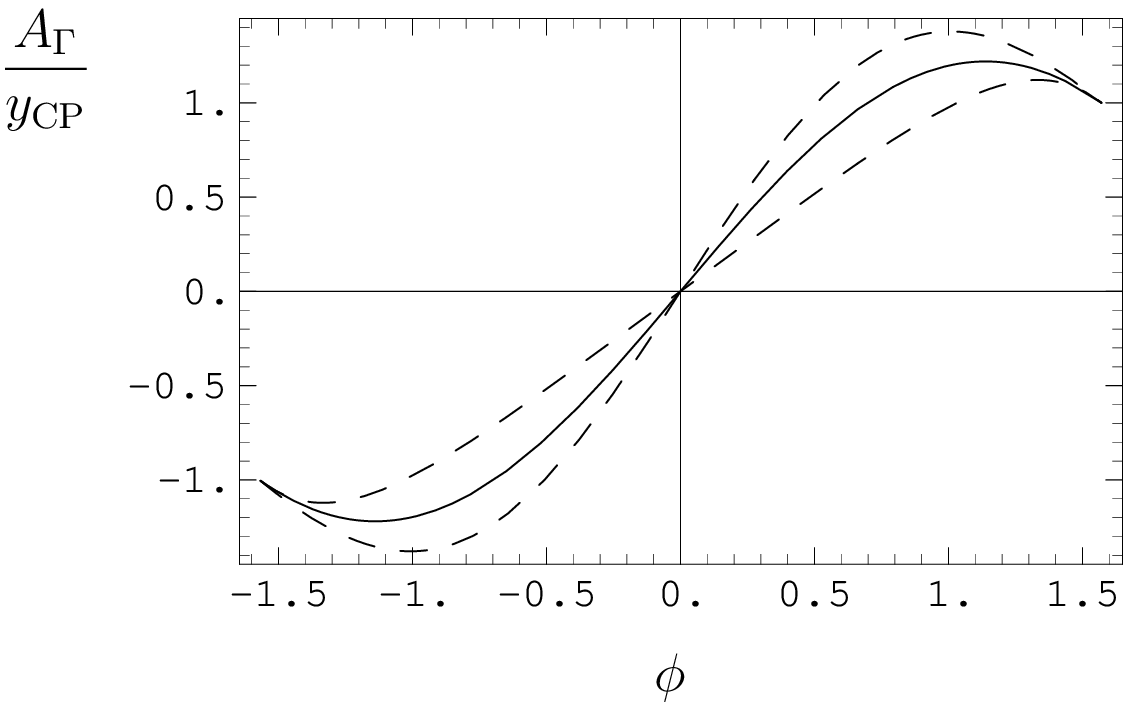}$$
\vskip-20pt
\caption[]{\small Left: $y_{\rm CP}/y$ as function of $\phi$, for $x/y=1.2$
  (solid line) and $x/y=\{0.6,1.8\}$ (dashed lines), see
  Eq.~(\ref{new}). Right:
  $A_\Gamma/y_{\rm CP}$ as function of $\phi$.}\label{fig4}
\end{figure}

Two more CP-sensitive observables related to $D^0\to K^+ K^-$ have
been measured by the Belle collaboration \cite{Belle}:
\begin{eqnarray}
y_{\rm CP} & = &\frac{1}{2\Gamma}\, [\Gamma(D^0\to K^+K^-) + \Gamma(\bar 
D^0\to K^+K^-)] -1\nonumber\\
& = & \frac{1}{2}\,\left(\left|\frac{q}{p}\right| +
\left|\frac{p}{q}\right| \right) y \cos\phi + 
\frac{1}{2}\,\left(\left|\frac{q}{p}\right| -
\left|\frac{p}{q}\right| \right) x \sin\phi,\\
A_\Gamma & = & \frac{1}{2\Gamma}\,  [\Gamma(D^0\to K^+K^-) -\Gamma(\bar 
D^0\to K^+K^-)] -1\nonumber\\
& = & \frac{1}{2}\,\left(\left|\frac{q}{p}\right| -
\left|\frac{p}{q}\right| \right) y \cos\phi + 
\frac{1}{2}\,\left(\left|\frac{q}{p}\right| +
\left|\frac{p}{q}\right| \right) x \sin\phi.
\end{eqnarray}
The present experimental value of $y_{\rm CP}$ is given in (\ref{8a}), that for
$A_\Gamma$ is $(0.01\pm 0.30({\rm stat}) \pm 0.15({\rm syst}))\times
10^{-2}$. Again, we can study the dependence of these observables on
$\phi$. In Fig.~\ref{fig4}a we plot the ratio 
$y_{\rm CP}/y$, which is a function
of $x/y$ and $\phi$, in dependence on $\phi$. As it turns out, this
quantity is far less sensitive to $\phi$ than $y'_+/y'_-$, the reason
being that its deviation from $1$ is only a second-order effect in $\phi$:
\begin{equation}
y_{\rm CP} = y \left\{ 1 + \phi^2\,\frac{x^4+x^2
  y^2-y^4}{2(x^2+y^2)^2} + O(\phi^4)\right\}.
\end{equation}
Hence, unless the experimental accuracy is dramatically increased, and
because the results on $y'_+/y'_-$ and $y/y_{\rm SM}$ already exclude
a large CP-odd phase $\phi\approx \pm \pi/2$, it
is safe to interpret $y_{\rm CP}$ as measurement of $y$.
In Fig.~\ref{fig4}b we plot the quantity $A_\Gamma/y_{\rm CP}$. Also
here there is a distinctive dependence on $\phi$, with
$A_\Gamma/y\propto \phi$ for small $\phi$, but the effect is
less dramatic than that in $y'_+/y'_-$.

In conclusion, we find that the experimental results on $D$ mixing
reported by BaBar and Belle already exclude extreme values of the
CP-odd phase $\phi$ close to $\pm \pi/2$. This follows from the
result for $y$, which is close to the top end of theoretical predictions
and can only be {\em reduced} by new physics, and from 
$y'_+/y'_-\sim 1$. While $y'_+/y'_--1$ vanishes in the limit of no
CP-violation, $y\sim\Delta\Gamma$ is a CP-conserving observable, which
demonstrates the usefulness of such quantities in constraining
CP-odd phases. 
Also $y_{\rm CP}$, $A_\Gamma$ and the ratio
$A_\Gamma/y_{\rm CP}$ can be useful in constraining $\phi$. 
As long as there is no
major breakthrough in theoretical predictions for $D$ mixing, which
are held back by the fact that the $D$ meson is at the same time too heavy and
too light for  current theoretical tools to get a proper grip on
the problem, the long-distance SM contributions to $x$ will completely
obscure any NP contributions and their detection. The observation of CP
violation, however, presents a theoretically clean way for NP to manifest
itself and it is to be hoped that in the near future, i.e.\ at the $B$
factories or the LHC, at least one of the plentiful opportunities for
NP to show up in CP violation~\cite{opportunities} will be realised.

\section*{Acknowledgments}

This work was supported in part by the EU networks
contract Nos.\ MRTN-CT-2006-035482, {\sc Flavianet}, and
MRTN-CT-2006-035505, {\sc Heptools}.

\end{document}